\documentstyle[aps,prl,epsfig]{revtex}

\begin{document}

\twocolumn[\hsize\textwidth\columnwidth\hsize\csname@twocolumnfalse\endcsname
\draft

\title{Fluctuation Effect in Superconductivity in an Electron Model with $d$-Wave Attaction}

\author{Xin-Zhong Yan}

\address{Institute of Physics, Chinese Academy of Sciences, P. O. Box 603, Beijing 100080, China} 

\date{\today}

\maketitle

\widetext
\begin{abstract}
On the basis of $t$-matrix approximation, we study the superconductivity in the tight-binding model with $d$-wave attraction. The low-lying collective modes are considered as the predominant long-range fluctuations in the system. The Green's function is obtained as an analytic solution to a cubic equation. The superconducting order parameter and the transition temperature are substantially reduced from the values of the mean-field theory. The calculated phase boundary of the superconductivity can reasonably describe the experiment results for the cuprates. The results for density of states and the pseudogap parameter are presented.
\end{abstract}

\vfill
\narrowtext

\vskip2pc] 

The fluctuation effect takes important role for describing the superconductivity in a quasi-two-dimensional superconductor with low carrier density [1]. The superconducting order parameter and thereby the transition temperature $T_c$ can be substantially suppressed from the values given by the mean-field theory (MFT). In most cases, the long-range pairing fluctuation is important as it prohibits the off-diagonal-long-range ordering in systems of dimensions $\leq$ 2 [2]. For the under doped high-$T_c$ cuprates (HTC), Emery and Kivelson attributed the suppression on $T_c$ to the long-range classical phase fluctuation [3]. Above $T_c$, hole pairing becomes locally without long-range phase coherence. On the other hand, since there are preformed pairs above $T_c$, the superconductivity below $T_c$ can be viewed as a consequence of Bose-Einstein condensation [4-6]. Along with the approach by Leggett [4], much effort has been devoted to investigation of the crossover from the weak-coupling BCS superconductivity to the Bose-Einstein condensation (BEC) of bound pairs [7-16]. Most of the work has been performed for the $s$-wave pairing because of its computational simplicity.

In this paper, we investigate the superconductivity in the tight-binding model with $d$-wave attraction. The low-lying collective modes of the boson fields above the condensate are treated as the predominant fluctuations. We present a self-consistent formalism for the Green's function based on the $t$-matrix approximation. The single particles and the collective modes are treated on the equal-footing manner. The new result is that below $T_c$, with the condensation taking place, the single pairs become moving collectively. Even at the ground state, there remains the zero-point motion. We calculate the phase boundary of superconductivity and compare the result with experiment data of HTC. We also study the pseudogap phenomena at $T_c$.

The Hamiltonian of the electron system is given by 
$$
H = \sum_{k\alpha}\xi_k c_{k\alpha }^{\dagger }c_{k\alpha }+
{\frac{1}{N}}\sum_{kk'q}v_{kk'}p^{\dagger }(k,q)p(k',q) \eqno(1)
$$
where $c_{k\alpha}^{\dagger }$ ($c_{k\alpha}$) is the creation (annihilation) operator for electrons with momentum-$k$ and spin-$\alpha $, $\xi_k = -2t(\cos k_x + \cos k_y) - 2t_z\cos k_z -\mu$ with $\mu$ the chemical potential, $N$ the total number of lattice sites, $v_{kk'} = -v\eta_k\eta_{k'}$ with $\eta_k = \cos k_x - \cos k_y$, and $p(k,q) = c_{-k+q/2\downarrow}c_{k+q/2\uparrow}$ is the pair operator. The differences between Eq. (1) and the
$t-J$ model are, (1) here only the $d$-wave attraction channel in the interaction is taken into account, (2) the prohibition of doubly occupation on the same site is released by assuming the hopping integrals $t$ and $t_z$ are proportional to the hole concentration $\delta$, e.g., $t = t_0\delta$ with $t_0$ a constant. For the quasi-two-dimensional system, $t_z/t \ll 1$ is supposed. This model has been adopted by a number investigators for studying the $d$-wave superconductivity as well as the pseudogap phenomena in cuprates [12-14,17].

In Nambu's representation, the Green's function $G(k,z_n)$ of the electrons is given by
$$
G(k,z_n) = [z_n - \xi_k\sigma_3 - \Sigma(k,z_n)]^{-1} \eqno(2)
$$
where $z_n = i\pi T(2n+1)$, $n$ is an integer, $T$ the temperature, and $\sigma$ the Pauli matrix.
Throughout this paper, we use the units in which $\hbar = k_B = 1$. To express the self-energy, firstly, we note that the off-diagonal part comes from averaged boson fields of momentum $q = 0$. In the superconducting state, $\langle p(k,0)\rangle$ is a macroscopic quantity as compared with any other pair fields else. Therefore, the predominant contributions are the static mean filed
$$
\Sigma_{12}(k,z_n) = {\frac {1}{N}}\sum_{k'}v_{kk'}\langle p(k',0)\rangle \equiv \Delta_k \eqno(3)
$$
and its complex conjugate $\Sigma_{21}(k,z_n)$. For our uniform system, we suppose $\Delta_k$ is real. The quantity $\Delta_k \equiv \Delta\eta_k$ should be differentiated from that of the MFT since the fluctuation effect is under consideration in the present Green's function. Secondly, we take into account of the pair fluctuation terms $q \ne 0$ of the interaction in the diagonal part. By the $t$-matrix approximation, which is the infinite sum of ladder diagrams, the diagonal part of the self-energy is given by
$$
\Sigma_{\mu\mu}(k,z_n) = -{\frac{T}{N}}\sum_{qm}\eta^2_{k-q/2}G_{\bar\mu\bar\mu}(k-q,z_n-Z_m)W_{\mu\mu}(q,Z_m) \eqno(4)
$$
where $Z_m =i2\pi Tm$, $m$ is an integer, $\mu$ = 1, 2 with $\bar 1 = 2$ and $\bar 2 = 1$, and the $t$ matrix $W(q,Z_m)$ is given by
$$
W(q,Z_m) = v^2[1 + v\chi(q,Z_m)]^{-1}\chi(q,Z_m) \eqno(5)
$$
where $W$ and $\chi$ are $2\times 2$ matrices, with the elements of $\chi$ defined by
$$
\chi_{\mu\nu}(q,Z_m) = {\frac{T}{N}}\sum_{kn}\eta^2_kG_{\mu\nu}(k+q/2,z_n+Z_m) G_{\bar\nu\bar\mu}(k-q/2,z_n). \eqno(6)
$$
The chemical potential $\mu$ is determined by
$$
{\frac{2T}{N}}\sum_{kn}G_{11}(k,z_n)e^{z_n\eta} = 1-\delta \eqno(7)
$$
where $\eta$ is an infinitesimal positive number. These equations (2)-(7) form the closed system that self-consistently determines the Green's function and the $t$ matrix.

Two points about the present formalism need to be emphasized. Firstly, the existence of Goldstone mode in the superconducting state requires that the pair susceptibility $\chi(q,Z_m)$ should satisfy the condition [18]
$$
\det|1 + v\chi(0,0)| = 0. \eqno(8)
$$
This equation is exactly consistent with Eq. (3). Any improper treatment of off-diagonal self-energy leads to violation of this consistency. At $T_c$, equation (8) reduces to the Thouless criterion [19]. Secondly, because of the divergence of $W_{\mu\mu}(q,Z_m)$ at $q = 0$ and $Z_m = 0$, the diagonal self-energy takes into account of the predominant fluctuation effect.

It is a tremendous task to numerically solve Eqs. (2)- (7) because many multi-dimensional integrals over momentum and the summation over Matsubara's frequency need to be computed in each iteration. However, since $W(q,Z_m)$ is strongly peaked with a divergence about $q = 0$ and $Z_m = 0$, the diagonal self-energy can be approximately given by [14,20]
$$
\Sigma_{\mu\mu}(k,z_n) \approx -\eta^2_kG_{\bar\mu\bar\mu}(k,z_n){\frac{T}{N}}\sum_{qm}{'}W_{\mu\mu}(q,Z_m)e^{\alpha_{\mu}Z_m\eta} \eqno(9)
$$
where $\sum{'}$ means the summation over $q$ runs a small $q$ region, and the convergent factor $e^{\alpha_{\mu}Z_m\eta}$ with $\alpha_1 = 1$ and $\alpha_2 = -1$ has been introduced. This convergent factor comes from the fact that the Green's function $G_{\bar\mu\bar\mu}(k-q,z_n-Z_m)$ in the summation in Eq. (4) is connected with the effective interaction $W_{\mu\mu}(q,Z_m)$. The summations over $q$ and $m$ in Eq. (9) give rise to a constant
$$
\Gamma^2 = -{\frac{T}{N}}\sum_{qm}{'}W_{\mu\mu}(q,Z_m)e^{\alpha_{\mu}Z_m\eta}. \eqno(10)
$$
At small $q$ and $Z_m$, the $t$ matrix can be approximated by the collective modes. At $T < T_c$, the collective modes are sound-like waves with energy $\Omega_q \propto q$, while at $T_c$, $\Omega_q \propto q^2$, the excitations are single pairs. The constant $\Gamma^2$ is essentially a measure of the density of these uncondensed pairs. $\Gamma$ is named as pseudogap parameter since at $T_c$ there remains a gap in the density of states (DOS) at the Fermi energy. The $q$-integral in Eq. (10) is over a cylinder region in the momentum space. Since $\Omega_q$ depends weakly on the out-plane wave number $q_z$, the integral over $q_z$ can be taken in the range $(-\pi,\pi)$. The cutoff $q_c$ for the in-plane wave number is determined such that the largest in-plane energy $\Omega_{q_{c}} = 2\sqrt{\Delta^2+\Gamma^2}$ since the collective mode is meaningful only within the gap.

Now, note that two equations from the diagonal parts of Eq.(2) with $\Sigma_{\mu\mu}$ given by Eq. (9) form the closed system for determining the diagonal Green's functions. By eliminating one of them, we can obtain a cubic equation for $G_0$ or $G_3$. By introducing a function $y(k,z_n)$, the Green's function is obtained as 
$$
G(k,z_n) = [z_n + 3\Delta_k\sigma_1/(1+y) + \xi_k\sigma_3](2-y)/3\Gamma^2_k. \eqno(11)
$$
where $\Gamma_k = \Gamma\eta_k$. The function $y(k,z_n)$ is a real root to the cubic equation
$$
y^3 - 3P y - 2Q =0 \eqno(12)
$$
where $P = 1+3(\Gamma^2_k-\Delta^2_k)/(\xi^2_k-z_n^2)$ and $Q = 1+{\frac{9}{2}}(\Gamma^2_k+2\Delta^2_k)/(\xi^2_k-z_n^2)$. The explicit form of $y(k,z_n)$ reads
$$
y = \left\{
\begin{array}{ll}
\sqrt[3]{Q+\sqrt{D}}+\sqrt[3]{Q-\sqrt{D}},&\mbox{$D>0$}\cr
2\sqrt{P}\cos (\varphi/3),&\mbox{$D<0$}
\end{array}\right.
\eqno(13)
$$
where $D = Q^2-P^3$, and $\varphi = \arccos(Q/\sqrt{P^3})$. The boundary condition $y \to 2+3\Gamma^2_k/(\xi^2_k-z_n^2)$ at $|z_n| \to \infty$ is useful for analytic contituation to real frequency $z_n \to \omega+i\eta$. 

For describing the cuprates, we take $v/2t_0 \simeq 0.1$ and $t_z/t \simeq 0.01$ [13]. For $\rm La_2CuO$, $v \simeq 0.13$ eV has been determined by experiments [21]. Therefore, our choice of $v/2t_0$ corresponds to $t_0 \simeq 0.65$ eV, which is consistent with estimates from experiment data [13,22]. The small quantity $t_z/t$ describes the interlayer weak coupling and gives rise to a $z$-freedom energy in $\Omega_q$. This weak coupling prevents the summation over $q$ in Eq. (10) from a logarithm divergence at the $q = 0$ limit and ensures a finite transition temperature $T_c$. 

The result for $T_c$ as a function of hole concentration $\delta$ (solid line) is plotted in Fig. 1. The MFT result (with the same scale as the present theory) and the experiment data [23] are also shown for comparison. The maximum transition temperature $T_{c,Max} \approx$ 108 K obtained by the present theory appears at a certain $\delta$ between 0.125 and 0.15. The improvement of the present theory to the MFT is significantly. The phase boundary of superconductivity given by the present theory is close to the experiment results. In the optimally to over doped region, the present theory fits the experiment data very well. Contrary to the MFT in the under doped region, $T_c$ of the present calculation increases reasonably with $\delta$. Such a behavior of $T_c$ reflects the feature of BEC [19].

There is still obvious discrepancy between the present theory and the experiment at small $\delta$. This may date from the crude treatment of the short-range pair correlations. Local pairing without long-range phase coherence is not fully taken into account in the present model. Besides this, the short-range antiferromagnetic coupling is not correctly counted in. To describe the antiferromegnetism in cuprates at very small $\delta$, one needs to restart with the $t-J$ model. In addition, long-range Coulomb interaction may also take effect if there is no adequate screening. To what extent all these factors affect the superconductivity may be a subject for future study. 

The reduction of $T_c$ from its MFT value stems from that a part of the pairs occupy collective modes. As mentioned above, the density of these pairs is measured by $\Gamma^2$. In Fig. 2, the result for the pseudogap parameter $\Gamma$ at $T_c$ is compared with the MFT $\Delta_{MF}(T_c)$ as well as the pseudogap energy $E_g$ determined by experiments [24]. The experimental observations indicate that $E_g$ slightly depends on $T$ above $T_c$ [25]. We here plot the result for $E_g$ to view the overall magnitudes only. The parameter $\Gamma$ similarly as $\Delta_{MF}(T_c)$ decreases monotonously with $\delta$. The larger ratio $\Gamma/\Delta_{MF}(T_c)$ at smaller $\delta$ implies the larger occupation of the uncondensed pairs. This is consistent with the more significant reduction of $T_c$ in the under doped region.

In Fig. 3, we show the numerical results for $\Delta$ and $\Gamma$ as functions of $T$ at $\delta = 0.1$. The superconducting gap opens below $T_c$ and reaches its maximum at $T = 0$. The parameter $\Gamma$ decreases with $T$ decreasing and remains a finite value at $T = 0$. At the ground state, the fluctuation effect comes mainly from the zero-point motion (quantum fluctuation) of the pairs. Generally speaking, in a quasi-two-dimensional system, the order parameter of the broken-symmetry state can be considerably suppressed by the quantum fluctuation. This can be confirmed also by the perturbation calculations [26-28]. The Chicago group [12-14] has considered the BCS-BEC crossover problem. By their approach, the uncondensed pairs below $T_c$ are single pairs with energy $\Omega_q \propto q^2$, and completely condensation takes place at $T = 0$. By the present treatment, however, the BEC from the single pairs begins to occur at $T_c$, while below $T_c$, with the condensation taking place, the uncondensed pairs become moving collectively. At $T = 0$, the zero-point motion prevents pairs from fully BEC.

The pairing fluctuation effect can be clearly reflected in the DOS of single quasiparticles, $\rho(E) = -2t\sum_{k}{\rm Im} G_{11}(k,E+i\eta)/\pi N$. A direct effect is the pairing fluctuation introduces lifetime to the single quasiparticles. Shown in Fig. 4 are the results for the DOS at $\delta = 0.1$ and $T/T_c$ = 0.2 and 1. $\rho(E)$ depends on $E$ linearly at small $E$. This is in the character of $d$-wave gap. In contrast to the well-known MFT, the peaks in the DOS are broadened even at $T < T_c$ due to the collective modes. The width scales with $\Gamma$. At $T_c$, $\rho(E)$ shows still the existence of the $d$-wave gap with magnitude about $2\Gamma$. That is the pseudogap. To understand better of the psedogap, we consider the spectral function at $T_c$, $A(k,E) = \sqrt{(\xi_k^2+4\Gamma_k^2-E^2)/(E^2-\xi_k^2)}(E+\xi_k)/2\pi\Gamma^2_k$ [20] which is nonzero only for $E^2-4\Gamma^2_k < \xi_k^2 < E^2$, with the noninteracting delta function peak becoming a square root singularity. For $E^2-4\Gamma^2_k < 0$, a part of spectral weight is shifted away. This results in the formation of a pseudogap in the DOS. 

In summary, we have investigated the superconductivity in the tight-binding model with $d$-wave attraction. On the basis of $t$-matrix approximation, we have developed the self-consistent formalism for the Green's function. The analytic Green's function is given by Eq. (13). The low-lying collective modes are treated as the predominant long-range pairing fluctuation in the self-energy. The pairing fluctuation results in lifetime effect for the single particles below $T_c$ and a pseudogap in the density of states at $T_c$. The transition temperature is substantially suppressed from its mean-field value. The phase boundary of superconductivity given by the present theory is close to the experiment results for the cuprates.

This work was supported by NSF of China (GN. 10174092) and Department of Science and Technology of China (G1999064509).

\begin{figure}[tbp]
\centerline{\epsfig{file=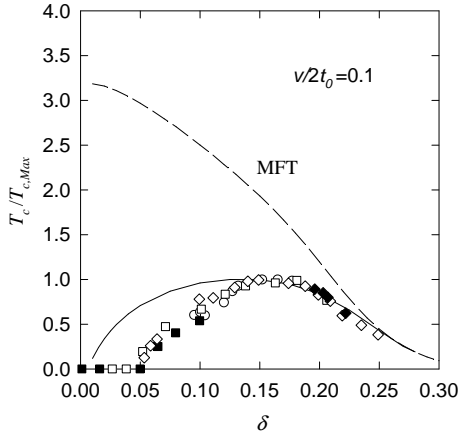,width=6 cm}}
\vskip 2mm
\caption{$T_c$ as a function of $\delta$. The solid and dashed lines represent the results of present approach and the MFT, respectively. The symbols indicate the experiment data for cuprates [23]: $\rm Y_{1-x}Ca_xBa_2Cu_3O_6$ (solid squares), $\rm Y_{0.9}Ca_{0.1}Ba_2Cu_3O_{7-y}$ (open squares), $\rm La_{2-x}Sr_xCuO_4$ (open diamonds), $\rm Y_{1-x}Ca_xBa_2Cu_3O_{6.96}$ (solid diamonds), and $\rm YBa_2Cu_3O_{7-y}$ (open circles).}
\end{figure}

\begin{figure}[tbp]
\centerline{\epsfig{file=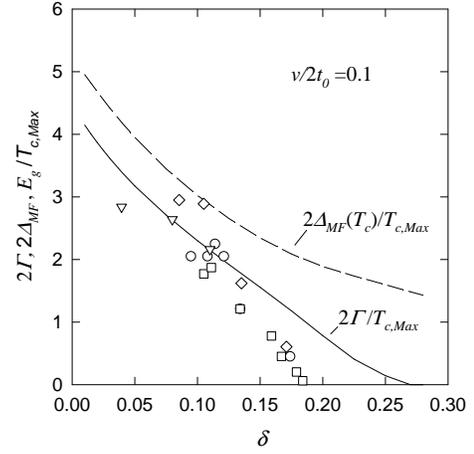,width=6 cm}}
\vskip 2mm
\caption{$\Gamma$ (solid line) and $\Delta_{MF}$ (dashed line, by the MFT) as functions of $\delta$ at $T_c$. The symbols represent the pseudogap energy $E_g$ determined by experiments for cuprates [24].}
\end{figure}

\begin{figure}[tbp]
\centerline{\epsfig{file=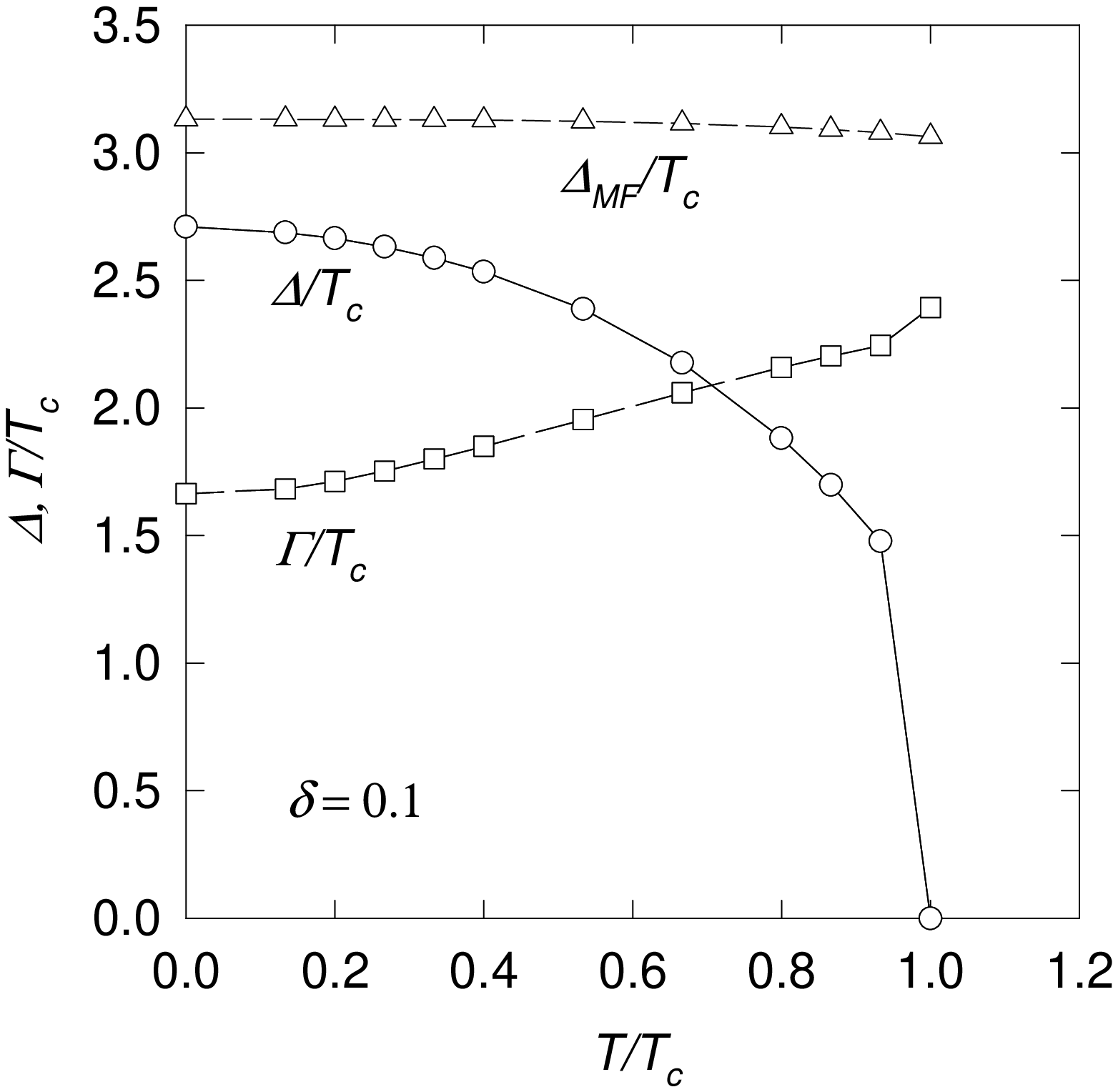,width=6 cm}}
\vskip 2mm
\caption{$\Delta$ and $\Gamma$ as functions of $T$ at $\delta = 0.1$. $\Delta_{MF}$ given by the MFT is also plotted for comparison.}
\end{figure}

\begin{figure}[tbp]
\centerline{\epsfig{file=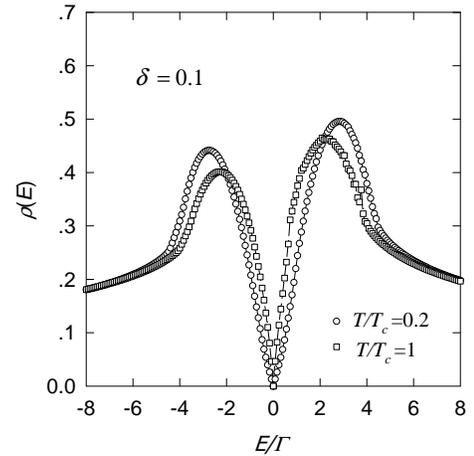,width=6 cm}}
\vskip 2mm
\caption{Density of states $\rho(E)$ at $T/T_c = 0.2$ and 1. The hole concentration is $\delta = 0.1$. }
\end{figure}

\end{document}